\documentstyle[graphicx]{article}

\def\abstract#1{\vskip 7mm
        \begin{center}{\large Abstract}\par \smallskip
                \begin{minipage}[c]{12cm}
                        \small #1
                \end{minipage}
        \end{center}
}
\def\title#1{\begin{center}{\Large\bf #1}\end{center}}
\def\author#1{\vskip 5mm \begin{center}{#1}\end{center}}
\def\address#1{\begin{center}{\it #1}\end{center}}
\def\beq{\begin{equation}}
\def\eeq{\end{equation}}
\def\C{{\cal C}}
\def\B{{\cal B}}
\def\R{{\cal R}}
\def\d{{\delta}}
\def\E{{\cal E}}
\def\H{{\cal E}}
\def\A{{\cal A}}

\makeatletter
\@ifundefined{lesssim}{\def\lesssim{\mathrel{\mathpalette\vereq<}}}{}
\@ifundefined{gtrsim}{\def\gtrsim{\mathrel{\mathpalette\vereq>}}}{}
\def\vereq#1#2{\lower3pt\vbox{\baselineskip1.5pt \lineskip1.5pt
\ialign{$\m@th#1\hfill##\hfil$\crcr#2\crcr\sim\crcr}}}
\makeatother

\begin{document}

\title{%
  Gravitation and cosmology in a brane-universe
  \smallskip \\
}
\author{%
  David Langlois\footnote{E-mail:langlois@iap.fr}
}
\address{%
Institut d'Astrophysique de Paris, \\
98bis Boulevard Arago, 75014 Paris, France
}

\abstract{Recent theoretical developments have generated a strong interest
in the ``brane-world''  picture,
which assumes that ordinary matter is trapped in a three-dimensional
submanifold, usually called brane, embedded in a higher dimensional
space. The purpose of this  review is to introduce some  basic results 
concerning gravity in these models and then to present various 
aspects of  the cosmology 
in a brane-universe. }

\section{Introduction}
The idea that our world may contain hidden extra dimensions is 
rather old since one can trace  this idea, in the modern context of 
general relativity, 
back to the beginning of the twentieth century, with the 
pioneering works of Kaluza and Klein, trying to reinterpret 
electromagnetism as a geometrical effect from a fifth dimension. 
The idea of extra-dimensions was revived more recently with the advent 
of string theory as the most promising avenue for conciling 
gravity and quantum field theory. 
In order to get a consistent 
theory at the quantum level, ten spacetime dimensions are needed 
in superstring theories (eleven in M-theory), 
which means that six dimensions must be 
somehow hidden to four-dimensional observers such as us.

The simplest way to hide  extra dimensions is to assume that they 
are flat, compact with a  radius  sufficiently small to be 
unobservable. Consider for example the case of one extra-dimension
described by the  coordinate $y$ and compactified via the identification 
$$
y\rightarrow y+ 2\pi R,
$$
$R$ being the ``radius'' of the extra-dimension.
Any matter field, for example a scalar field, depends on both the ordinary 
spacetime coordinates $x^\mu$ and the extra-coordinate $y$. It  
  can be Fourier expanded  along the extra-dimension
so that 
\beq
\phi(x^\mu,y)=\sum_{p=-\infty}^\infty e^{ipy/R} \phi_p(x^\mu).
\eeq
The corresponding
 Fourier modes $\phi_p$ 
are  designated as Kaluza-Klein modes, and each of them 
can be seen as a four-dimensional scalar field satisfying the 
four-dimensional Klein-Gordon equation with the effective squared 
mass
\beq
M^2_p=m^2+{p^2\over R^2}.
\eeq
A simple way to identify an extra-dimension would thus be to detect 
the characteristic spectrum of the Kaluza-Klein modes.  
To do this one needs enough  energy to excite a least the first Kaluza-Klein
modes and the non observation of Kaluza-Klein modes can be interpreted as 
meaning that the size of the extra-dimension is smaller than the 
inverse of the energy scale probed by the experiment.
Present constraints from colliders thus imply
$$
R\lesssim 1 \ ({\rm TeV})^{-1}.
$$

Let us now turn to gravity. The natural way to extend Einstein gravity 
to higher dimensions is to start from  the  Einstein-Hilbert action defined 
in a generalized  spacetime with, say, $n$ extra-dimensions:
\beq
S_{grav}=\int d^4x\, d^ny \,  {R\over 2\kappa^2},
\eeq
where $R$ is the scalar curvature in the $(4+n)$-dimensional spacetime.
Variation of the action including the matter part leads 
 to the generalized Einstein equations  which read
\beq
G_{AB}\equiv R_{AB}-{1\over 2}R\  g_{AB}=\kappa^2 T_{AB}.
\label{einstein}
\eeq
This equation has exactly the same form as the familiar one, with the 
difference that  
all tensors are now  $(4+n)$-dimensional tensors.

In the static weak field regime (in a flat $(4+n)$-dimensional spacetime) 
Einstein equations imply  as usual the Poisson equation. However,
the solution of Poisson's equation depends on the  number 
of space dimensions, and the general form of the Newtonian potential is 
\beq
\phi_N(r)\propto {G_{(4+n)}\over  r^{n+1}},
\eeq
where the generalized Newton's constant $G_{(4+n)}$ is proportional to the 
gravitational coupling $\kappa^2$ introduced above in the Einstein equations
(\ref{einstein}).

This means that, a priori, the presence of extra-dimensions 
implies  that gravity is  modified. 
The simplest way to recover the familiar gravity law is once more 
to compactify the extra-dimensions. The 
resulting gravity is 
\begin{itemize}
\item the $(4+n)$ dimensional gravity on scales smaller than 
the  radius $R$ of compactification (assumed here 
to be the same in all extra-dimensions for simplicity)
\item the usual $4$-dimensional gravity on scales much larger than $R$, 
with the Newton's constant given by 
\beq
G_{(4)}\sim G_{(4+n)}/R^n,
\eeq
or, equivalently, expressed in terms of the Planck mass, 
\beq
M_{(4)}^2\sim M_{(4+n)}^{2+n} R^n.
\label{vol_extra}
\eeq
\end{itemize}
This result can be understood is the following way. 
In a compactified space,  the gravitational field  induced 
by a point mass $m$  can be computed by unwrapping the extra-space and by 
summing  the contributions of  all the images of the 
true mass $m$. 
At small distances with respect to  $R$, the 
influence of the image masses can be ignored and one gets 
$(4+n)$ dimensional gravity. By contrast, on scales much larger 
than $R$, all image masses contribute to the gravitational field and they 
can be assimilated to a continuous massive ``line'' 
with constant ``linear'' mass density. Applying then Gauss' law to a cylinder 
surrounding the massive line yields the usual gravitational force with 
the above gravitational coupling. 

As a consequence, like in particle physics, an upper constraint on the 
compactification radius can be deduced from the absence 
 of any observed  deviation from ordinary Newton's law. 
The present experimental  constraints yield (see e.g. \cite{grav_exp}) 
\beq
R \lesssim  0.2 \, {\rm mm}.
\eeq

The latest developments in the models with extra-dimensions  come from 
the realization that the observational constraint on the size 
of extra-dimensions from gravity experiments 
is much weaker than that from accelerator experiments. 
This suggests  the idea to decouple the extra-dimensions
``felt'' by ordinary particles from the extra-dimensions ``felt'' by gravity.
Concretely, this can be realized by invoking a mechanism that confines
fields of the particle physics Standard Model to a subspace with 
three spatial dimensions, called three-brane, within a higher dimensional 
space where gravity  lives.  

The purpose of this review, far from being exhaustive, 
is to present  the basic results concerning gravity  and cosmology in the 
brane scenarios where the self-gravity of the brane is taken into account, 
and to illustrate some more advanced aspects like cosmological 
perturbations or brane collisions. 
Complementary information can be found in two recent reviews, one by 
R. Maartens \cite{maartens}
 and the other by V. Rubakov \cite{rubakov}, the latter 
more  focused on the particle physics aspects.

\section{Braneworld models}

The braneworld scenarios have attracted much attention only 
recently but one can find in the literature  a few precursor works 
\cite{precursors} considering our four-dimensional universe as a subspace of 
a larger spacetime. 
A wider interest in braneworlds has developed  when this idea has 
emerged in the context of superstring theories and M-theory. 
An important step is the 
Horava-Witten supergravity \cite{hw}, 
which is supposed to describe the effective 
low energy theory in the strong coupling limit of heterotic 
$E_8\times E_8$ superstrings.  This theory is based on an eleven-dimensional 
spacetime with an orbifold $S_1/Z_2$ eleventh dimension, with 11-dimensional
supergravity in the bulk and a gauge group $E_8$ on each of the 
10-dimensional boundaries corresponding to the worldsheets of 
9-branes located at the two fixed points along the eleventh dimension. 
After compactification of six dimensions, one finds a five-dimensional 
spacetime with two four-dimensional boundaries where matter is supposed to 
be localized. 

At the phenomenological level, Arkani-Hamed, Dimopoulos and Dvali (ADD) 
\cite{add}  then  suggested the more radical step 
 to assume that the fundamental Planck mass is of the order of 
the TeV so as to solve the famous hierarchy problem in particle physics, 
i.e. why there are so many orders of magnitude between the electroweak scale 
and the Planck scale. 
In these models 
the observed Planck scale  seems huge simply because the volume of the extra 
dimensions is large, according to the  formula (\ref{vol_extra}). 
The ADD proposal has had a tremendous impact, especially 
  in the particle physics community.

In the general relativity and cosmology communities, more interest was 
suscitated by the later proposal of Randall and Sundrum \cite{rs99a,rs99b}. 
By contrast with the ADD models, their specificities are the following:
\begin{itemize}
\item there is only one  extra-dimension

\item the bulk  spacetime is not flat but curved: it is 
a portion of  Anti-de Sitter
 (AdS) with a negative cosmological constant 
\beq
\Lambda\equiv -{6\over \ell^2}\equiv -6\mu^2
\eeq
where $\ell$ has the dimension of length, and $\mu$ the dimension of mass.  

\item  there is a tension, $\sigma$,  in the brane(s), which is (are) supposed
to be $Z_2$-symmetric like in the Horava-Witten model. 
\end{itemize}
The metric can then be written in the form
\beq
ds^2=a^2(y)\eta_{\mu\nu} dx^\mu dx^\nu +dy^2,
\eeq
where $\eta_{\mu\nu}$ is the usual Minkowski metric and the 
warping factor is given by
\beq
a(y)=e^{-|y|/\ell}.
\eeq
There are in fact two models due to Randall and Sundrum, with essentially  
 the same framework but which differ in the r\^ole  
 assigned to  the positive tension brane:
\begin{itemize}
\item RS1: two branes are bounding the AdS portion, one positive tension 
brane at $y=0$ and one negative tension brane, the latter corresponding 
to our accessible world \cite{rs99a}. 
This model suggests a solution to the hierarchy 
problem, by showing that TeV energy scales on the hidden brane 
(the TeV brane) correspond to $M_P$ energy scales in our brane 
(the Planck brane), due to the exponential warping factor.

\item RS2: in this model \cite{rs99b}, 
our world corresponds to the positive tension 
brane which is located at $y=0$. The negative tension brane is now
facultative. The goal of this model is not to solve 
the hierarchy problem, but to show that an infinite extra dimension 
can lead to usual four-dimensional gravity, as explained below. 
 \end{itemize} 
The tension of the branes is not a free parameter for the above setup to be
valid. They must be fine-tuned with the AdS cosmological constant so as 
to satisfy the condition
\beq
{\kappa^2\over 6}|\sigma|={1\over \ell}.
\label{rs}
\eeq

From now on, I will consider only the second model RS2 and show 
how ordinary gravity is approximately recovered.  In order to 
study weak gravity, the usual method  is 
 to consider tensor-like  linear  perturbations of the metric. 
Denoting these fluctuations  $h_{AB}$ so that 
\beq
g_{AB}=\bar{g}_{AB}+h_{AB},
\eeq
and imposing  the gauge requirements $h_{Ay}=0$, $h_{\mu\ ,\nu}^{\nu}=0$ and
$h_\mu^\mu=0$, 
the linearized
Einstein equations yield the following wave equation for the 
gravitons
\beq
\left[a^{-2}\Box^{(4)}+\partial_y^2-{4\over \ell^2}
+{4\over \ell}\delta(y)\right]h_{\mu\nu}=0, 
\label{RS_wave_eq}
\eeq
where the delta function takes into account the presence of the brane.
This equation is separable and the general solution can be written as the 
superposition of solutions  of the form
\beq
h_{\mu\nu}(x^\lambda, y)=u_m(y)e^{ik_\lambda x^\lambda}\epsilon_{\mu\nu},
\eeq
where $m$ (satisfying $k_\lambda k^\lambda=-m^2$)
can be interpreted as  the four-dimensional mass 
and 
$u_m(y)$ satisfies an ordinary differential equation, which is deduced 
from (\ref{RS_wave_eq}) by replacing the d'Alembertian 
operator with $m^2$.
Following Randall and Sundrum, it is convenient to rewrite this equation 
in a Schr\"odinger-like form,
\beq
-{d^2\psi_m\over dz^2}+V(z)\ \psi_m=m^2\psi_m
\eeq
where $\psi_m=a^{-1/2}u_m$ and $z=sgn(y)\ell(\exp(|y|/\ell)-1)$. The 
potential $V(z)$ is ``volcano''--shaped, its expression being  given by
\beq
V(z)={15\over 4(|z|+\ell)^2}-{3\over \ell}\delta(z).
\eeq
The delta function part of the potential leads to the existence of 
a zero mode, with the functional dependence
$$
u_m\propto a^2,
$$
which is (exponentionally) localized near the brane and which reproduces
the usual four-dimensional graviton, the four-dimensional 
gravitational coupling being given by
\beq
8\pi G_{(4)}=\kappa^2/\ell.
\label{G4}
\eeq
 In addition to this zero mode, 
there is a whole continuum of massive graviton modes, which induce 
some corrections to the usual gravitational law, although significant only 
on scales of the order of the AdS lengthscale $\ell$ and below. The 
resulting gravitational potential
is of the form
\beq
V(r)={G_{(4)}\over r}\left(1+ \alpha {\ell^2\over r^2}\right).
\eeq
The coefficient $\alpha=1$, given initially by RS, was corrected to 
$\alpha=2/3$ by  a more 
detailed analysis \cite{gt},  which  
 considered  explicitely the coupling of graviton to matter in the brane.

Finally, let us mention that a  rewriting   of the 
Einstein equations, in  the 
Randall-Sundrum type models, leads  to  effective four-dimensional 
Einstein equations, which can be written in the form \cite{sms} 
\beq
{}^{(4)}G_{\mu\nu}=8\pi G_{(4)} \tau_{\mu\nu}+\kappa^4\Pi_{\mu\nu} -E_{\mu\nu}.
\eeq
where $\tau_{\mu\nu}$ is the brane energy-momentum tensor 
(not including the tension $\sigma$),  
$\Pi_{\mu\nu}$ is quadratic in the brane energy momentum tensor, 
\beq
\Pi_{\mu\nu}=-{1\over 4}\tau_{\mu\sigma}\tau^{\sigma}_\nu
+{1\over 12}\tau\tau_{\mu\nu}+{1\over 8} g_{\mu\nu}\left(\tau_{\sigma\rho}
\tau^{\sigma\rho}-{1\over 3}\tau^2\right).
\eeq
 and 
 $E_{\mu\nu}$ is the projection of the five-dimensional Weyl tensor
\beq
E_{\mu\nu}={}^{(5)}C^{A}_{\ BCD}n_A n^C
g_\mu^B g_\nu^D,
\eeq
 $n^A$ being the unit vector normal to the brane. 
Although very nice, one must be aware that the above equation is only 
a rewriting of the five-dimensional Einstein's equations using the 
junction conditions, and in practical problems, the full system remains 
to be solved. In particular one must be careful with the interpretation
of the gravitational coupling  to brane matter, even in the linear case, 
because $E_{\mu\nu}$ will in general depend on the matter $\tau_{\mu\nu}$.

In a two-brane model, in contrast with the single brane model, what is 
obtained is Brans-Dicke type gravity with the radion, i.e. the interbrane
separation playing the r\^ole of the Brans-Dicke scalar field \cite{gt}.
The model RS1 leads to  a Brans-Dicke  gravity 
which is incompatible with observations: this model can 
 be saved only by introducing a more complicated setting, for example
a scalar field in the bulk with a potential and couplings to the branes, 
which provides  an  stabilization mechanism for the radion
\cite{gw}.

\section{Homogeneous cosmology in a brane-universe}

Let us now turn to cosmology. 
The main motivation for exploring cosmology in models with extra-dimensions 
is that the potentially new effects could arise   significantly  
 only at very high energies, i.e. in the very early universe, 
and leave some relic imprints which could be  tested today via
 cosmological observations.
Before discussing the potentially rich but very difficult question of 
cosmological perturbations in the next section, 
one must first describe homogeneous cosmology, following \cite{bdl99}
and \cite{bdel99} (see also \cite{ftw99}).

Let us thus consider a five-dimensional spacetime with three-dimensional
isotropy and homogeneity, which contains a three-brane representing
our universe. It is convenient, but not necessary, 
 to work in a Gaussian normal coordinate
system based on  our brane-universe. Due to the spacetime 
symmetries, the metric is then of the form
\beq
ds^2=- n(t,y)^2 dt^2+a(t,y)^2 \delta_{ij}dx^idx^j+dy^2,
\label{metric}
\eeq
where we have assumed that our brane-universe is spatially flat (but 
this can be generalized very easily to hyperbolic or elliptic spaces).
In these coordinates, our brane-universe is always located at $y=0$.

The energy-momentum tensor can be decomposed into a bulk 
energy-momentum tensor and a brane energy-momentum tensor, the latter 
being of the form
\beq
 T^A_{\, B}= S^A_{\, B}\delta (y)= \{\rho_b, p_b, p_b, p_b, 0\}\delta (y),
\eeq
where the delta function expresses the confinement of matter 
in the brane. $\rho_b$ and $P_b$ are respectively 
the total energy density and pressure in the brane and depend only
on time.
For simplicity, we neglect the bulk energy-momentum 
tensor  but  allow for the presence of a cosmological 
constant in the bulk, $\Lambda$,  
 so that the five-dimensional Einstein equations read
\beq
G_{AB}+\Lambda g_{AB}=\kappa^2 T_{AB}.
\eeq
Because of the distributional nature of the energy-momentum tensor, 
one way to solve the Einstein equations is to solve them first in the 
bulk  and then apply the junction 
conditions \cite{israel} for the metric
 at $y=0$. 
According to the junction conditions,   the metric must be 
continuous and  the jump of the extrinsic curvature tensor 
$K_{AB}$  (related
to the derivatives of the metric with respect to $y$) depends on 
the distributional energy-momentum tensor, 
 \beq
\left[K^A_{\, B}
-K\delta ^A_{\, B}\right]=\kappa^2 S^A_{\, B},
\label{israel}
\eeq
where the brackets here denote the jump at the 
brane, i.e. $[Q]=Q_{\{y=0^+\}}-Q_{\{y=0^-\}}$, and the extrinsic curvature 
tensor is defined by 
\beq
K_{AB}=h_{A}^C\nabla_C n_B,
\eeq
$n^A$ being the unit vector normal to the brane.
As before, one can  add the extra assumption that the brane is 
mirror symmetric 
so that the jump in the extrinsic curvature is twice its value on one 
side (see \cite{nonZ2} for references where this is not assumed). 
Substituting  the ansatz metric (\ref{metric}) in (\ref{israel}), one ends up 
with the two junction conditions:
\beq
\left({n'\over n}\right)_{0^+}={\kappa^2\over 6}\left(3p_b+2\rho_b\right),
\qquad
\left({a'\over a}\right)_{0^+}=-{\kappa^2\over 6}\rho_b.
\label{junction}
\eeq

Going back to the bulk Einstein equations (their explicit form can be found 
in e.g. \cite{bdel99}), one can solve the $(t-y)$ component to get
\beq
\dot a(t,y)=\alpha(t)n(t,y), 
\eeq
and the integration of the component $(t-t)$ with respect to $y$ and 
of the component $(y-y)$ with respect to time, yields the  first integral
\beq
(aa')^2-\alpha^2 a^2+{\Lambda\over 6} a^4+\C=0,
\eeq
where $\C$ is an integration constant. 
When one evaluates this first integral at $y=0$, i.e. in our brane-universe, 
substituting the junction conditions given above in (\ref{junction}), 
one ends up with the following equation 
\beq
H_0^2\equiv {\dot a_0^2\over a_0^2}={\kappa^4\over 36}\rho_b^2+{\Lambda\over 6}
+{\C\over a^4}.
\label{fried}
\eeq
where the subscript `$0$' means evaluation at $y=0$.
This equation is analogous to the  (first)
Friedmann equation, since it relates the Hubble parameter to the 
energy density, but it is different from the usual Friedmann equation
[$H^2=(8\pi G/3)\rho$]. 
 Its most  remarkable feature  is that the energy density of the brane enters 
quadratically on the right hand side in contrast with the standard 
four-dimensional Friedmann equation where the energy density enters 
linearly. 
As for  
the  energy conservation equation it 
is unchanged in this five-dimensional setup
and still reads
\beq
\dot\rho_b+3H(\rho_b+p_b)=0.
\eeq

In the simplest case where $\Lambda=0$ and $\C=0$, 
one can easily solve the above cosmological equations for 
a perfect fluid with an equation of state $p_b=w\rho_b$ and $w$ constant.
One finds that the evolution of the scale factor is given by
\beq
a_0(t)\propto t^{1\over 3(1+w)}.
\eeq
In the most interesting cases for cosmology, radiation and pressureless 
matter, one finds respectively  
$a\sim t^{1/4}$  (instead 
of the usual $a\sim t^{1/2}$) and $a\sim t^{1/3}$  (instead 
of  $a\sim t^{2/3}$).
Such behaviour  is problematic because it cannot be reconciled 
with nucleosynthesis. Indeed, the nucleosynthesis scenario depends crucially 
on both  the microphysical  reaction rates
 and the expansion rate of the universe. And changing in a drastic way 
the evolution of the scale factor between nucleosynthesis and now 
 modifies dramatically the predictions for the light element abundances.

The above Friedmann law with the $\rho_b^2$ term, but without the 
bulk cosmological constant (and without the $\C$ term) was first derived in 
\cite{bdl99}, just before  Randall and Sundrum proposed their models. 
In fact, the unusual Friedmann law 
 can be related to a gravity which should be  five-dimensional
rather than four-dimensional. With the obtention in RS2 of a five-dimensional 
model yielding a four-dimensional gravity, one could 
expect that the corresponding 
cosmology should  be compatible with the usual cosmology \cite{cosmors}.
In fact, it is clear from  (\ref{fried}) that a Minkowski brane with a 
tension, as in the RS2 model, requires the presence of a negative 
cosmological constant in the bulk to compensate the squared tension term 
and get $H=0$.

If one wants to go beyond a Minkowski geometry  and consider 
non trivial   cosmology in the brane, one must then assume 
that the total energy density 
in the brane, $\rho_b$, consists of two parts, 
\beq
\rho_b=\sigma+\rho,
\eeq
the tension $\sigma$, constant in time, and the usual cosmological energy
density $\rho$. 
Substituting this decomposition into (\ref{fried}), one obtains 
\beq
H^2= \left({\kappa^4\over 36}\sigma^2-\mu^2\right)
+{\kappa^4\over 18}\sigma\rho
+{\kappa^4\over 36}\rho^2+{\C\over a^4}.
\eeq
If one fine-tunes the brane tension and the bulk cosmological cosmological 
constant as in (\ref{rs}), 
the first term on the right hand side vanishes.
The second term then becomes the dominant  term if $\rho$ is small enough and
{\it one thus recovers the usual Friedmann equation at low energy}, 
with the identification
\beq
8\pi G= {\kappa^4\over 6}\sigma,
\label{newton}
\eeq
which is exactly the relation obtained in RS2 by combining (\ref{rs}) and 
(\ref{G4}). 

The third term on the right hand side, quadratic in the energy density, 
provides a {\it high-energy correction} to the Friedmann equation 
which becomes significant when the value of the energy density approaches 
the value of the tension $\sigma$ and dominates  at higher 
energy densities. In the very high energy regime, $\rho\gg \sigma$, one 
thus recovers the unconventional behaviour analysed before since the 
bulk cosmological constant becomes negligible.
It is in fact not difficult to obtain explicit solutions for the scale 
factor, which interpolate between the low energy regime and the 
high energy regime.

Finally, the last term on  the right hand side behaves like radiation and 
arises from the integration constant $\C$. This 
constant $\C$ is quite analogous to the Schwarzschild mass and it is  
related to the bulk Weyl tensor, which vanishes when $\C=0$. In a 
cosmological context, this term is constrained to be small enough 
at the time of nucleosynthesis in order to satisfy the constraints on the 
number of extra light degrees of freedom. In the matter era, this term 
then redshifts quickly and would be in principle negligible today.

In the present section,  we have so far considered only  the metric 
in the brane. The metric  outside the brane can be also determined explictly
\cite{bdel99}.
In the special  case $\C=0$, the metric has a much simpler form and 
its components are given by   
\begin{eqnarray}
a(t,y)&=& a_0(t)\left(\cosh\mu y-\eta \sinh\mu|y|\right)\\
n(t,y)&=& \cosh\mu y-\tilde\eta \sinh\mu|y|
\label{bulk_metric}
\end{eqnarray}
where
\beq
\eta=1+{\rho\over\sigma}, \qquad \tilde\eta=\eta+{\dot\eta\over H_0}
\eeq
and we have chosen the time $t$ corresponding to the cosmic time in the 
brane. 
In the RS2 limit, $\rho=0$, i.e. $\rho_b=\sigma$, which 
implies  $\eta=\tilde\eta=1$ and one recovers
$a(t,y)=a_0\exp(-\mu|y|)$.

In summary, we have obtained a cosmological model, based on a braneworld
scenario, which appears to be 
compatible with current observations a low enough
energies. Let us now quantify the constraints on the parameters of the 
model in order to ensure this  compatibility with observations. 
As mentioned above an essential  constraint  comes 
from nucleosynthesis: the evolution of the universe since
nucleosynthesis must be  approximately 
the same as in usual cosmology. This is the case if 
the energy scale associated with the tension is higher than 
the nucleosynthesis
energy scale, i.e.
\beq
M_c\equiv\sigma^{1/4} \gtrsim 1 \ {\rm MeV}.
\eeq
Combining this with (\ref{newton}) this implies for the fundamental mass scale
(defined by $\kappa^2=M^{-3}$)  
\beq
M \gtrsim 10^4 \ {\rm GeV}.
\eeq
There is however another constraint, which is not of cosmological nature:
the requirement to recover ordinary gravity down to scales of the 
submillimeter order. This implies
\beq
\ell \lesssim  10^{-1} \ {\rm mm},
\eeq
which yields the constraint
\beq
 M\gtrsim  10^8 \ {\rm GeV}.
\eeq
Therefore the most stringent constraint comes, not from cosmology, but
from gravity experiments in this particular model.
So far, we have thus been able to build a model, which reproduces all 
qualitative and quantitative features of ordinary cosmology in the 
domains that have been tested by observations. 
The obvious next question is whether this will still hold for a more 
realistic cosmology that includes  perturbations  from homogeneity, and more 
interestingly, whether  
brane cosmology is capable of 
providing  predictions that  deviate from usual cosmology and 
which might tested in the  future. This is still an open question 
today.

\section{Brane cosmological perturbations}
Endowed with a viable homogeneous scenario, 
one would like to explore the much richer domain 
of cosmological perturbations and investigate whether brane cosmology 
leads to new effects that could be tested in the forthcoming 
cosmological observations, in particular of 
the anisotropies of the Cosmic Microwave Background.

Brane cosmological perturbations is a difficult subject and although there 
are now many published works  on this question (see e.g. 
\cite{mukohyama,kis,maartens01,l00a,bdbl,koyama,l00b,ddk00,lmsw00}), 
no observational  signature has yet been  
 predicted. 
Below I will summarize some results concerning two differents aspects 
of perturbations. The first aspect deals with the evolution of 
scalar type perturbations on the brane, the second aspect with the 
production of gravitational waves from quantum fluctuations during a 
de Sitter phase in the brane.

Let us first  discuss scalar type cosmological perturbations in brane 
cosmology. 
Choosing a GN coordinate system the ``scalarly'' perturbed metric can 
be written 
\beq
ds^2=-n^2 (1+2 A)dt^2+2n^2 \partial_i B dt dx^i 
+a^2\left[ (1+2C)\d_{ij}+2\partial_i \partial_j E\right]dx^i dx^j
+dy^2,
\eeq
where the perturbations turn out to coincide exactly with 
 the standard scalar cosmological perturbations 
since we are in the GN gauge. One can 
find other gauge choices in the literature. 
 Using the compact notation 
$h_\alpha=\{A,B,C,E\}$ ($\alpha=1,\dots, 4$),
the linearized Einstein equations 
\beq
\delta G_{AB}+\Lambda \d g_{AB}=\kappa^2 \d T_{AB}
\eeq
yield, {\it in the bulk}, expressions of the form
\begin{eqnarray}
\delta G^{(5)}_{00}&=&  \delta 
G^{st}_{00}+
[h_\alpha, h'_\alpha,h''_\alpha] =-\Lambda\delta g_{00}\\
\delta G^{(5)}_{ij} &=& \delta G^{st}_{ij}+
[h_\alpha, h'_\alpha,h''_\alpha]=-\Lambda\delta g_{ij}\\
\delta G^{(5)}_{0i} &=&  \delta G^{st}_{0i}+
[h_\alpha, h'_\alpha,h''_\alpha] =-\Lambda\delta g_{0i} \\
\delta G^{(5)}_{05} &=&  [h_\alpha, h'_\alpha]=0  \\
\delta G^{(5)}_{5i} &=&  [h_\alpha, h'_\alpha] =0\\
\delta G^{(5)}_{55} &=&  [h'_\alpha, h''_\alpha]=0
\end{eqnarray}
where the brackets represent linear combinations of the perturbations 
and their derivatives.
Brane matter enters only in the junction conditions, which at the linear 
level relate the first derivatives (with respect to $y$) of the 
metric perturbations $h'_\alpha$ to brane matter perturbations 
$\delta\rho$, $\delta P$, $v$, $\pi$. One can then substitute these 
relations back into the perturbed Einstein equations. 
$\delta G^{(5)}_{05}=0$ then yields the usual perturbed  
energy  conservation equation, whereas 
$\delta G^{(5)}_{i5}=0$ yields the perturbed Euler equation.
The other equations yield equations of motion for the perturbations 
where one recognizes the usual equations of motion in ordinary cosmology, 
but with  two  types of corrections:
\begin{itemize}
\item modification of the  homogeneous background coefficients due to the 
additional terms in the Friedmann equation.   
These  corrections are  negligible in the low
energy  regime 
$\rho\ll\sigma$.
For long wavelength (larger than the Hubble scale) perturbations, 
one can thus obtain  a transfer coefficient, $T=5/6$, characterizing the 
high/low energy   transition, i.e. 
\beq
\Phi_{\rho\ll\sigma}={5\over 6}\Phi_{\rho\gg\sigma}.
\eeq
\item presence of source terms in the  equations. 
These terms come from the bulk perturbations and cannot be determined solely 
from the evolution inside the brane. To determine them, one must solve 
the full problem in the bulk (which also means to specify some initial 
conditions in the bulk). From the four-dimensial point of view, these 
terms from the fifth dimension appear like external source terms and their 
impact is formally similar to that of ``active seeds'', which have been 
studied in the context of topological defects. 
\end{itemize}

Let us turn now to another facet of the brane cosmological perturbations: 
their origin. In standard cosmology, the main mechanism for producing 
the cosmological perturbations is inflation. One can thus try to 
generalize this mechanim to the context of brane cosmology. Brane 
inflation generated by a scalar field confined to the brane  has been 
investigated in \cite{mwbh99}. The spectrum of gravitational waves generated 
in such a scenario is however more subtle to compute because gravitational 
waves have an extension in the fifth dimension. It has been first 
computed in \cite{lmw00} and confirmed (and extended) in a different 
approach \cite{grs01}. 

To compute the production of gravitational waves, one 
can approximate slow-roll brane inflation by a succession of de Sitter phases.
The metric for a de Sitter brane (see also \cite{kaloper99}) 
corresponds to a particular case of 
(\ref{bulk_metric}) with $\eta=\tilde\eta$
 and can 
be written as 
\beq
a(t,y)=a_0(t) \A(y), \quad n=\A(y),
\eeq
with 
\beq
\A(y)=   \cosh\mu y-\left(1+{\rho\over\sigma}\right) \sinh\mu|y|.
\eeq
The gravitational waves appear in a perturbed metric of the form 
\beq
ds^2=-n^2 dt^2
+a^2\left[ \d_{ij}+E_{ij}^{TT}\right]dx^i dx^j
+dy^2,
\eeq
where the  `TT' stands for transverse traceless. 
Decomposing $E_{ij}^{TT}$ in Fourier modes of the 
form 
\beq
E_{ij}=E \ e^{i \vec k. \vec x}\  e_{ij},
\eeq
one gets a wave equation, which reads
\beq
\ddot{E}+3H_0\dot{E}+{k^2\over a_o^2}E=\A^2 E''+4\A \A'E'\,.
\label{E}
\eeq
This equation is separable, and one can look for solutions 
 $E=\varphi_m(t) \E_m(y)$, where the time-dependent part must 
satisfy
\beq
\ddot{\varphi}_m +3H_0\dot{\varphi}_m+\left[ m^2+{k^2\over
a_0^2}\right] \varphi_m =0\,, \label{varphieom}
\eeq
and the $y$-dependent part satisfies 
\beq
\H_m''+4{\A'\over\A}\H_m'+ {m^2\over \A^2}\H_m = 0\,.
\label{Heom}
\eeq
Like in the Minkowski case, the latter equation can be reformulated as 
a Schroedinger type equation, 
\beq
\label{SE}
 {d^2\Psi_m\over dz^2} - V(z)\Psi_m =-m^2 \Psi_m \,,
\end{equation}
 after introducing  the conformal 
coordinate $z=\int dy/\A(y)$ and defining $\Psi_m\equiv \A^{3/2}\H_m$.
The potential is given by 
\beq
V(z)= {15H_0^2 \over 4\sinh^2(H_0 z)} +
{\textstyle{9\over4}}H_0^2
- 3\mu\left[1+{\rho\over\sigma}\right] \delta(z-z_{\rm b}) \,.
\eeq
The non-zero value of the Hubble parameter signals the presence of a gap 
between the zero mode and the continuum of Kaluza-Klein modes, as noticed
earlier by \cite{gs99}. 

The zero mode corresponds simply to 
\beq
\H_0= C_1\equiv \sqrt{\mu}\,\,F\!\left({H_0/\mu}\right)\,,
\eeq
where, imposing the normalization $2 \int_{z_{\rm b}}^\infty |\Psi_0^2| dz=1$,
 the constant $C_1$ has been expressed in terms of $H_0$ via 
the function 
\begin{equation}
F\!\left(x\right) =\left\{ \sqrt{1+x^2} - x^2 \ln \left[ {1\over
x}+\sqrt{1+{1\over x^2}} \right] \right\}^{\!\!-1/2}
\!.\label{deff}
\end{equation}
Asymptotically, $F\simeq 1$ at low energies, i.e. $H_0\ll \mu$, 
and  $F\simeq \sqrt{3H_0/(2\mu)}$ at high energies, i.e. $H_0\gg \mu$.
One can then evaluate the vacuum quantum fluctuations of the zero mode 
by using the standard  canonical quantization. To do this 
explicitly, one writes the five-dimensional action for gravity at 
second order in the perturbations. Keeping only the zero mode and integrating
over the fifth dimension, one obtains  
\beq
S_{\rm g}= {1\over8\kappa^2}
\, \sum_{+,\times}\, \int d\eta\, d^3\vec{k}\,a_o^2\left[
\left({d\varphi_o \over
d\eta}\right)^2+k^2{\varphi_o}^2\right] \,, 
\eeq
This has the standard form for a massless graviton in
four-dimensional cosmology, apart from
the overall factor $1/8\kappa^2$ instead of $1/8\kappa_4^2$. It
follows that quantum fluctuations in each 
polarization, $\varphi_o$,
have an amplitude of $2\kappa (H_o/2\pi)$ on super-horizon scales. Quantum
fluctuations on the brane at $y=0$, where $E_o=C_1\varphi_0$, thus have the
typical amplitude
\beq
{1\over 2\kappa_4}\delta E_{\rm brane}=\left({H_0\over 2\pi}\right) F(H_0/\mu)
\eeq
At low energies, $F=1$ and 
one  recovers exactly the usual four-dimensional result 
but at higher energies the multiplicative factor $F$ provides an 
{\it enhancement of the gravitational wave spectrum amplitude 
with respect to the four-dimensional result}. However, comparing  
this with the amplitude for the scalar spectrum obtained in \cite{mwbh99},
one finds that, at high energies ($\rho\gg\sigma$), the {\it 
tensor over scalar 
ratio is in fact suppressed with respect to the four-dimensional ratio}.
An open question is how the gravitational waves will evolve during the 
subsequent cosmological phases, the radiation and matter eras.

\section{Collisions of branes}
The last topic I would like to discuss is of interest for scenarios
 with several branes which are allowed to collide. 
So far, I have focused on  cosmology for a single brane embedded in a 
five-dimensional AdS spacetime. For homogeneous cosmology, if one assumes 
the bulk to be empty, then the  derivation summarized above 
proves that the bulk and the other branes it might contain  can 
affect our brane-universe only {\it via the constant $\C$ of 
the Weyl radiation term}, where $\C$ is a constant in time. This 
property can be understood, more formally, as  a generalized 
Birkhoff theorem \cite{bcg}. 

Another brane can however have a dramatic influence when it collides with 
the initial brane. This possibility,  which could provide a new 
interpretation  of the Big-Bang in our brane-universe, has raised some
interest recently, in particular the ekpyrotic scenario 
\cite{kost} based on the five-dimensional reduction of Horava-Witten 
model \cite{low}, 
but other, simpler, models have also been proposed \cite{collisions}. 

With K. Maeda and D. Wands \cite{lmw01}, 
I have recently given a general analysis of the 
collision of n-branes in a $(2+n)$-dimensional empty spacetime 
with $n$-dimensional isotropy and homogeneity, i.e. 
branes separated   by patches of Sch-AdS spacetimes (allowing for different 
Schwarschild-type mass and cosmological constant in each region), 
with the metric
\beq 
ds^2=-f(R)dT^2+ {dR^2\over f(R)} +R^2d\Omega_n^2,
\label{metric_AdS} 
\eeq where the `orthogonal' metric $d\Omega_n^2$
does not depend on either $T$ or $R$. The well-known case of a
Schwarzschild-(anti)-de Sitter spacetime corresponds to
$f(R)=k-(\C/R^{n-1})\mp(R/\ell)^2$.

Contrarily to the coordinates of the previous section, a brane is no longer
at rest in this coordinate system. It can be described by its trajectory 
 $(T(\tau), R(\tau))$, where $\tau$ is
the proper time. An alternative way \cite{kraus} to obtain the generalized Friedmann 
equation (\ref{fried}) is simply to write the junction conditions 
at the location of the moving brane, with the  $Z_2$ symmetry assumption, 
by  noting  that the brane coordinate $R$ can be reinterpreted 
as the scale factor of the induced metric, as is clear from the metric 
(\ref{metric_AdS}). It can also been shown
explicitly  
that the expression for the metric (\ref{metric}) 
given above in  GN coordinates  can 
indeed be deduced from  (\ref{metric_AdS}) by an appropriate change of 
coordinates \cite{msm99}. 

To analyse the collision, it
is   very convenient to introduce an angle $\alpha$, which characterizes 
the motion of the brane with respect to the coordinate system 
(\ref{metric}),  defined 
by 
\beq
\label{alpha}
\alpha=\sinh^{-1}(\epsilon\dot R/\sqrt{f}),
\eeq
where $\epsilon=+1$ if $R$ decreases from ``left'' to ``right'',  $\epsilon=-1$
otherwise.
Considering   a collision involving a total number of $N$ branes, 
both ingoing and outgoing, thus separated by $N$ spacetime regions 
\begin{figure}[t]
\centering
 \includegraphics[width=100mm]{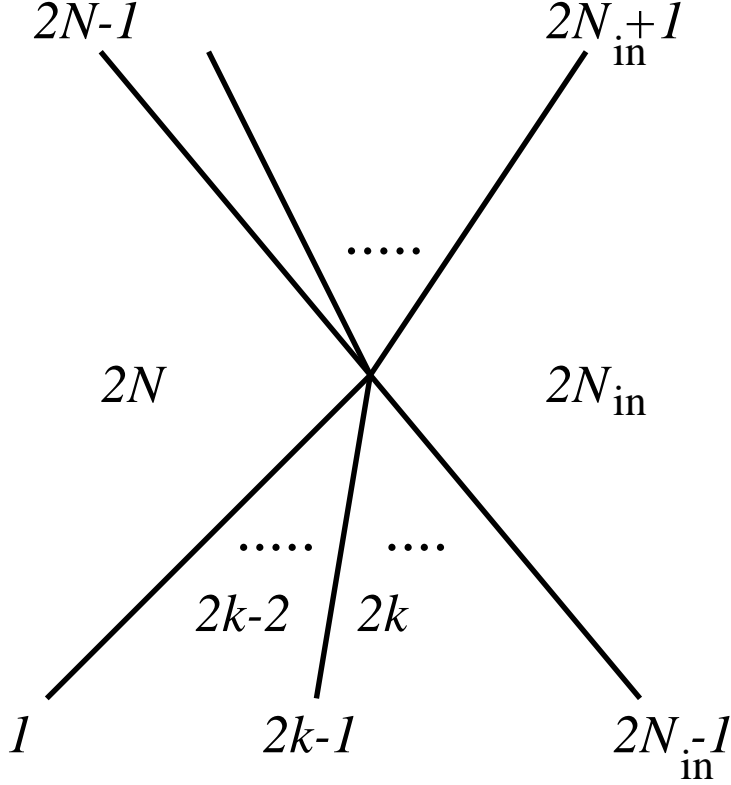}\\
\caption[Nbranes]{\label{Nbranes} Collision of $N_{\rm in}$ ingoing
  branes, yielding $N_{\rm out}=N-N_{\rm in}$ outgoing branes. Odd
  integers denote branes and even integers denote regions in between.}
\end{figure}
one can  label alternately branes
and regions by integers, starting from the leftmost ingoing brane and
going anticlockwise around the point of collision (see Figure).
 The branes will
thus be denoted by odd integers, $2k-1$ ($1\le k \le N$), and the
regions by even integers, $2k$ ($1\le k \le N$).  Let us introduce, as
before, the angle $\alpha_{2k-1|2k}$ which characterizes
the motion of
the brane $\B_{2k-1}$ with respect to the region $\R_{2k}$,
and which is defined by
\beq
\sinh\alpha_{2k-1|2k}={\epsilon_{2k}\dot R_{2k-1}\over \sqrt{f_{2k}}}.
\label{alpha_k}
\eeq
Conversely,  the motion of the region $\R_{2k}$
with respect to the brane by the Lorentz angle
$\alpha_{2k|2k-1}=-\alpha_{2k-1|2k}$.
It can be shown that the junction conditions for the branes can be written 
in the  form
\beq
\label{junction2} \tilde\rho_{2k-1} \equiv \pm{\kappa^2\over n}\rho_{2k-1} R=
\epsilon_{2k}\sqrt{f_{2k}}\exp{(\pm\alpha_{2k-1|2k})} 
 \ - \epsilon_{2k-2}\sqrt{f_{2k-2}}
\exp{(\mp\alpha_{2k-2|2k-1})} \label{rho},
\eeq
with the plus sign for ingoing branes ($1\le k\le N_{in}$), the minus 
sign for outgoing branes ($N_{in}+1\le k \le N$).
 An outgoing positive energy density brane
thus has the same sign as an ingoing negative energy density
brane.

The advantage of this formalism becomes obvious when one writes 
the {\it geometrical   consistency relation} that expresses the matching 
of  all branes and spacetime regions around the collision point. 
In terms of the angles defined above, it reads simply
\beq
\sum_{i=1}^{2N} \alpha_{i|i+1}=0. \label{collision}
\eeq
Moreover, introducing the generalized angles 
\beq
\label{relativeangle}
\alpha_{j|j'}=\sum_{i=j}^{j'-1}\alpha_{i|i+1},
\eeq
if $j<j'$, and $\alpha_{j'|j}=-\alpha_{j|j'}$,
the sum rule for angles (\ref{collision}) combined with the junction 
conditions (\ref{junction2}) leads to the laws of  energy conservation 
and momentum conservation. 
The energy conservation law reads
\beq
\sum_{k=1}^N\tilde\rho_{2k-1}\gamma_{j|2k-1}=0,
\eeq
where $\gamma_{j|j'}\equiv \cosh\alpha_{j|j'}$ corresponds to the Lorentz
factor between the brane/region $j$ and the brane/region $j'$ and can be
obtained, if $j$ and $j'$ are not adjacent, by combining all intermediary
Lorentz factors (this is simply using the velocity addition rule of
special relativity), or the relative angle formula~(\ref{relativeangle}).
The index $j$ corresponds to the reference frame with respect to which
the conservation rule is written.
Similarly, the {\it momentum conservation law} in the $j$-th reference frame
can be expressed in the form
\beq
\sum _{k=1}^N\tilde\rho_{2k-1}\gamma_{2k-1|j}\beta_{2k-1|j}=0,
\eeq
with $\gamma_{j|j'}\beta_{j|j'}\equiv \sinh\alpha_{j|j'}$.
One thus obtains, just from geometrical considerations, conservation laws
 relating  the brane energies densities and velocities before and after 
the collision point.
Our results apply to any collision of branes in vacuum, with the appropriate 
symmetries of homogeneity and isotropy. An interesting development would be 
to extend the analysis to branes with small perturbations and investigate 
whether one can find scenarios which can produce quasi-scale invariant 
adiabatic spectra, as seems required by current observations. 

\vskip 1cm 

\noindent
{\bf Acknowledgements}

I would like to thank the organizers of the 
11th Workshop on General Relativity and Gravitation
(Waseda University, Tokyo, Japan)
for a very interesting meeting. I would also like to acknowledge 
the  financial support of  the Yamada Science Foundation for my visit 
to Japan.

\end{document}